\documentclass[a4paper]{jpconf}
\usepackage[english]{babel}
\usepackage[utf8]{inputenc}
\usepackage[T1]{fontenc}
\usepackage{amsmath}
\usepackage{amssymb}
\usepackage[separate-uncertainty=true,
 range-units = single,
 range-phrase = --
]{siunitx}
\newcommand{\EeV}{\exa\electronvolt}

\usepackage{graphicx}

\begin{document}
\title{Cosmic Ray Physics with the LOFAR Radio Telescope}

\author{T~Winchen $^{1}$ and A~Bonardi $^{2}$, S~Buitink $^{1}$, A~Corstanje $^{2}$, H~Falcke $^{2,3,4}$, B~M~Hare $^{5}$, J~R~Hörandel $^{2,3}$, P~Mitra $^{1}$, K~Mulrey $^{1}$, A~Nelles $^{6}$, J~P~Rachen $^{2}$, L~Rossetto $^{2}$, P~Schellart $^{2,7}$, O~Scholten $^{5,8}$, S~ter~Veen $^{2,4}$, S~Thoudam $^{2,9}$, T~N~G~Trinh $^{5}$}

\address{$^1$ Astrophysical Institute, Vrije Universiteit Brussel, Pleinlaan 2, 1050 Brussels, Belgium,}
\address{$^2$ Department of Astrophysics/IMAPP, Radboud University, P.O. Box 9010, 6500 GL Nijmegen, The Netherlands,}
\address{$^3$ NIKHEF, Science Park Amsterdam, 1098 XG Amsterdam, The Netherlands,}
\address{$^4$ Netherlands Institute of Radio Astronomy (ASTRON), Postbus 2, 7990 AA Dwingeloo, The Netherlands,}
\address{$^5$ KVI-CART, University Groningen, P.O. Box 72, 9700 AB Groningen,\\}
\address{$^6$ Institut für Physik, Humboldt-Universität zu Berlin, Unter den Linden 6, 10099 Berlin, Germany,}
\address{$^7$ Department of Astrophysical Sciences, Princeton University, Princeton, NJ 08544, USA,}
\address{$^8$ Interuniversity Institute for High-Energy, Vrije Universiteit Brussel, Pleinlaan 2, 1050 Brussels, Belgium,}
\address{$^9$ Department of Physics and Electrical Engineering, Linn\'euniversitetet, 35195 V\"axj\"o, Sweden}

\ead{tobias.winchen@rwth-aachen.de}

\begin{abstract}
The LOFAR radio telescope is able to measure the radio emission from cosmic
ray induced air showers with hundreds of individual antennas.  This allows for
precision testing of the emission mechanisms for the radio signal as well as
determination of the depth of shower maximum $X_{\max}$, the shower observable
most sensitive to the mass of the primary cosmic ray, to better than
\SI{20}{\gram\per\square\centi\meter}.
	With a densely instrumented circular area of roughly \SI{320}{\square\meter},
LOFAR is targeting for cosmic ray astrophysics in the energy range
\SIrange{d16}{d18}{\electronvolt}.  In this contribution we give an overview of the
status, recent results, and future plans of cosmic ray detection with the LOFAR
radio telescope.
\end{abstract}

\section{Introduction}
The detection of cosmic rays via the radio technique has matured from
application in single engineering projects only to usage on large scales~\cite{Huege2016}. All major
experiments in the field nowadays use this technique at least complementary with
more traditional methods as surface detector arrays or fluorescence telescopes.
The radio technique allows for precision measurements of the depth of shower maximum
of air showers $X_{\max}$ and the energy deposit in the atmosphere.
Observation of cosmic rays with radio telescopes such as the Low Frequency
Array (LOFAR)~\cite{vanHaarlem2013} and potentially also the future Square Kilometer Array (SKA)~\cite{Huege2015} provide
observations of individual cosmic ray air showers with a high number of
antennas, enabling testing details of the radio emission mechanism and shower
development. High precision measurements of the cosmic ray mass composition in
the energy range below the ankle may be important to understand the expected
transition between galactic and extra-galactic cosmic rays~\cite{Thoudam2016a}.

LOFAR is the first digital radio telescope with more than 50 stations
distributed throughout Europe. Twenty-four of these stations are located in a
dense core in the Netherlands.  Five of the core stations are located within a
circle of \SI{320}{\meter} radius, the so called `superterp'.  Each core
stations consists of 96 `low-band antennas' (LBA) operating from
\SIrange{10}{90}{\mega\hertz} and 768 high-band antennas (HBA) operating from
\SIrange{110}{240}{\mega\hertz}. Both antennas are omnidirectional V-shaped
dipoles, but HBA are analog beamformed for astronomical observations
which diminish their usability for cosmic ray physics. In the following we will
thus discuss LBA only.

\section{Cosmic Ray Detection With LOFAR}
Cosmic ray detection with LOFAR currently relies on detecting the particle
cascade with an array of twenty scintillators, the LOfar Radboud air shower Array (LORA)~\cite{Thoudam2014}. These
scintillators are arranged in groups of four around the stations on the
superterp. If thirteen out of the twenty stations simultaneously detect
particles, indicating a high energy shower, \SI{2}{\milli\second} out of \SI{5}{\second} of buffered
antenna data are read out for further offline analysis.

The properties of the primary cosmic-rays are reconstructed by the comparison of the data with
dedicated simulations using the CORSIKA~\cite{Heck1998} and
CoREAS~\cite{Huege2013} software packages.  This yields a higher
resolution on $X_{\max}$ then using parametrizations of the radio profile in
ground plane~\cite{Nelles2015} and simplifies the evaluation of the detection
efficiency for the given shower parameters.

\begin{figure}[tb]
	\includegraphics[width=\textwidth]{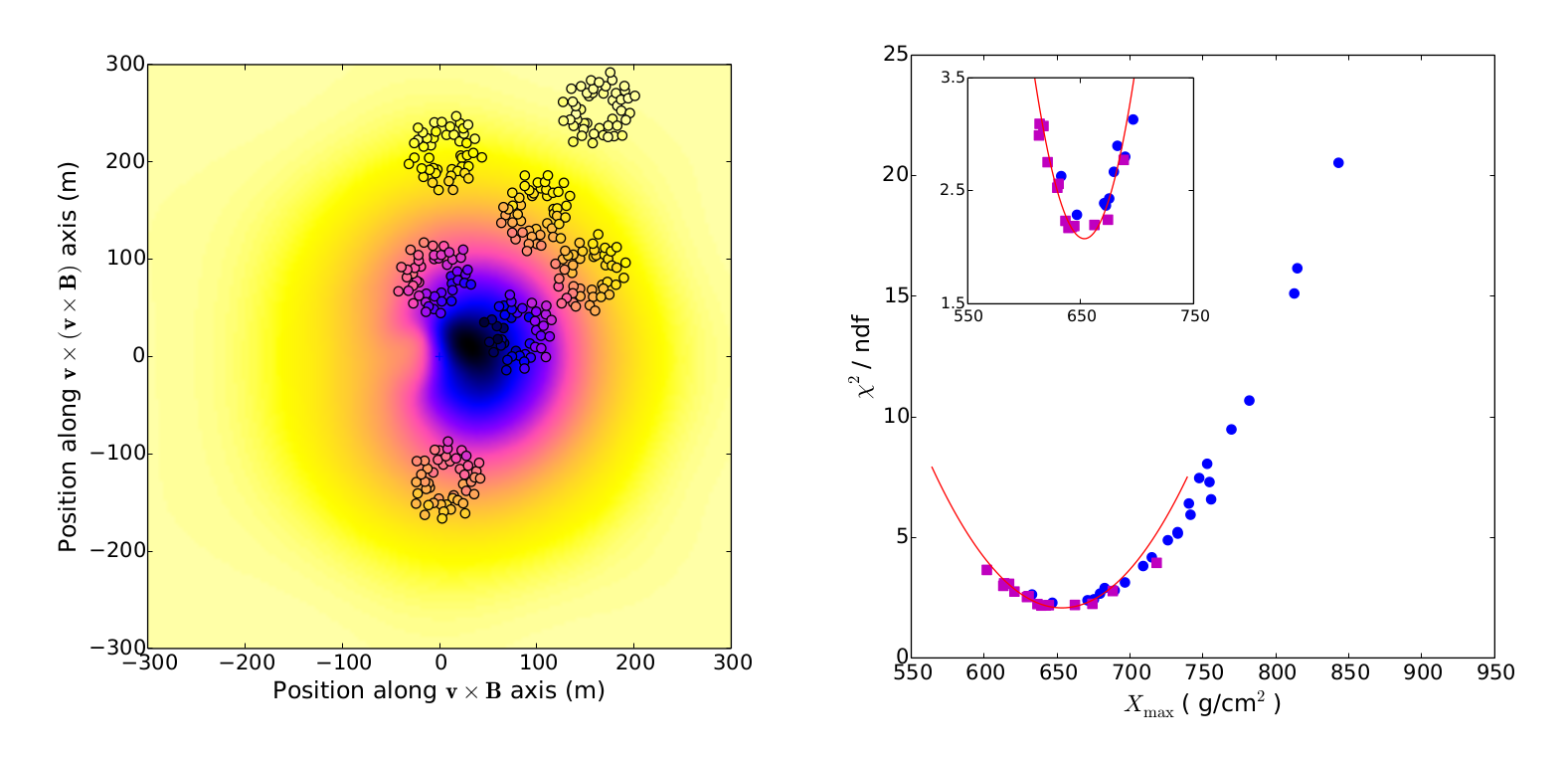}
	\caption{Event reconstruction with LOFAR. \textbf{(left)} Amplitude of the
	radio signal in the shower plane. Color in circles denote the observed
power in the LOFAR LBA, color in the background denotes the
signal from simulation of a shower with corresponding input parameters.
\textbf{(right)} Reduced chi-square from the comparison of simulated iron
(magenta squares) and proton (blue dots) showers with varying $X_{\max}$ with
the data. }
	\label{fig:XMaxReco}
\end{figure}
An example of a detected event and the reconstruction is shown in
Figure~\ref{fig:XMaxReco}. The left plot shows an example of a recorded event
together with the best fitting simulated shower. The plot on the right the
reduced chi-square as measure of the goodness of fit for all simulated showers.
The $X_{\max}$ of the individual shower is obtained from a fit of a parabola to
the simulated values.

With this procedure, LOFAR measures the cosmic ray mass composition in the
energy range \SIrange{d17}{d17.5}{\electronvolt}. The reconstruction
achieves an energy resolution of \SI{32}{\percent} and a statistical
uncertainty of \SI{16}{\gram\per\square\centi\meter} on $X_{\max}$. The total systematic uncertainty on the energy
calibration is \SI{27}{\percent} and on the depth of shower maximum
$^{+14}_{-10}$\;\si{\gram\per\square\centi\meter}.  Improvements of the
accuracy of the mass-composition data in the energy range covered by LOFAR may
be vital to discriminate between models for the origin  of galactic and
extragalactic  cosmic rays.

\section{Improvement of Systematic Uncertainties}

\subsection{Atmospheric Corrections}
\begin{figure}[tb]
	\centering
	\includegraphics[width=.6\textwidth]{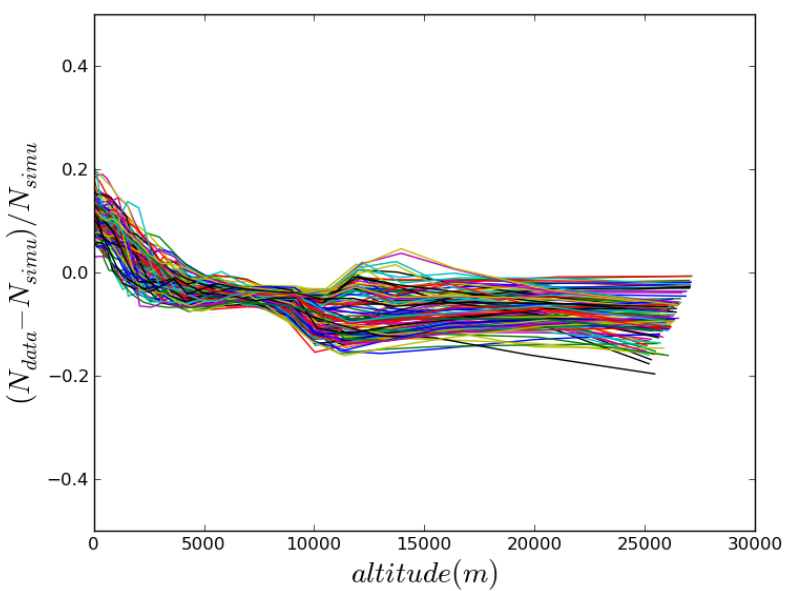}
	\caption{Variation of the refractive index along the shower profile for 100 different cosmic ray events observe with LOFAR.}
	\label{fig:atmospheric_Variation}
\end{figure}

The observed intensity distribution of the radio signal depends on the difference in propagation
time of particles and radio waves, and therefore on the refractivity profile
of the atmosphere at the time of the event. The refractivity of the atmosphere
depends on the density and humidity and  is thus expected to vary also on short
time scales~\cite{Corstanje2017a}.  In Figure~\ref{fig:atmospheric_Variation} the difference of the
refractive index at the time of 100 individual cosmic ray events to the US
standard atmosphere~\cite{NASA1976} is shown. At an altitude of
\SIrange{5}{8}{\kilo\meter} where the shower maximum is located, and thus the
bulk of radiation is emitted, the variation is of 3-5\%. This corresponds to an uncertainty in the inferred $X_{\max}$ of \SIrange{3.5}{11}{\gram\per\square\centi\meter}~\cite{Corstanje2017a}.

So far in LOFAR and, to our knowledge,
also in all  other cosmic ray experiments, standardized average atmospheric models
have been used for the interpretation of all showers. To account for the variation of the index of refraction, we started to use refractivity profiles obtained from atmospheric
data of the Global Data Assimilation Project (GDAS)~\cite{GDAS2004} in our simulations. For
this, we contributed a modification to the CORSIKA simulation software
 to
consistently use arbitrary atmospheric profiles in the simulation of the
particle cascade and radio signals.
The modification is included in CORSIKA version 7.6300 and following.
For this purpose  we fit the five layer
atmospheric model to the available data for the location of the observatory
and calculate the resulting index of refraction. The fit parameters describing the atmosphere layers, respectively
the tabualted index of refraction as a function of height, is then fed into the particle and radio simulation.
We developed a corresponding software
\texttt{gdastool} that manages data download, fit, and steering file
generation. This tool is also distributed alongside the
CORSIKA source code~\cite{Mitra2017}.

\subsection{Antenna Calibration}
To reconstruct amplitude and spectral shape of radio signals an accurate
calibration of the receiving system is imperative. However, at meter wavelength achieving the
accuracy in calibration necessary for cosmic ray physics with antenna systems
is challenging. No anechoic chambers are easily accessible
that can contain the antennas to be calibrated as well as the reference sources
in the far field region. Therefore even acquiring reference sources for in-situ
calibrations  that are well defined in the desired frequency range is
difficult. Alternatively to artificial reference sources, the ubiquitous
galactic background can be used as reference. Compared to artificial sources
the signal emitted by the galaxy is well understood in amplitude and shape.
However, the galaxy is not a point source so that calibrations using the galaxy
have to be based on a model for the directivity of the antenna.

\begin{figure}[tb]
	\includegraphics[width=\textwidth]{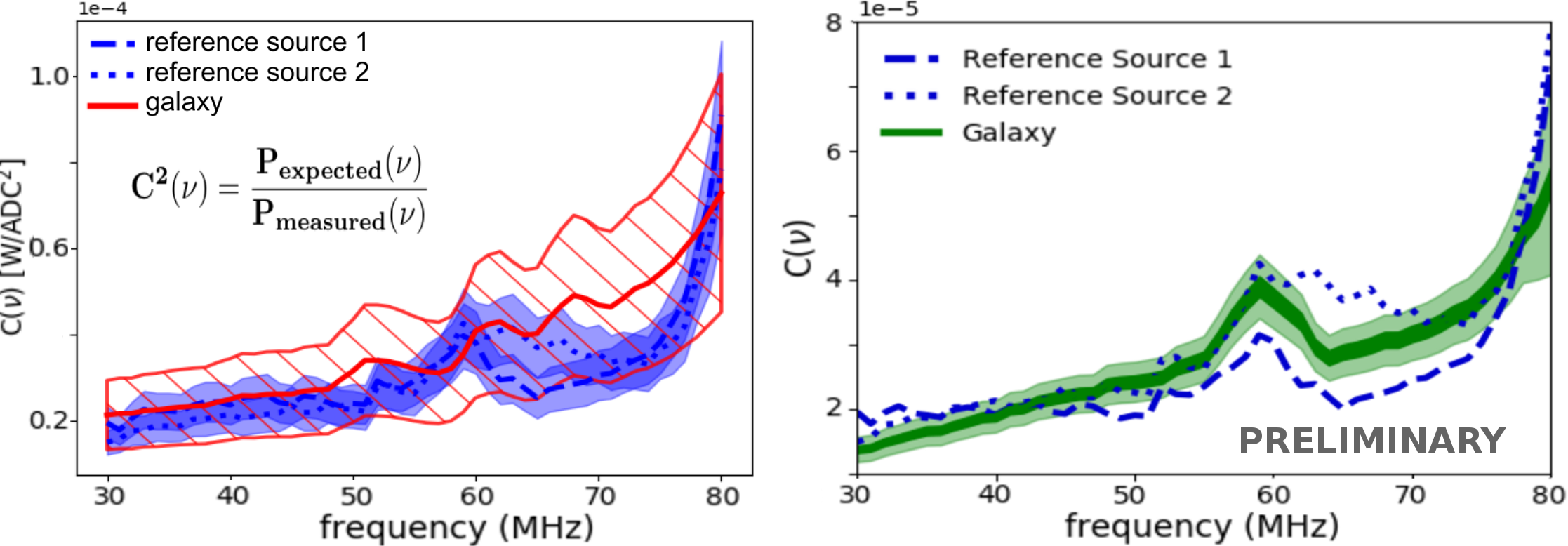}
	\caption{Antenna calibration factor as function of the frequency for galaxy calibration and reference source.
		\textbf{(left)} Without detailed model of the LOFAR signal chain and \textbf{(right)} with detailed model of the LOFAR signal chain.}
	\label{fig:antenna_calibration}
\end{figure}

\begin{figure}[tb]
	\centering
	\includegraphics[width=.6\textwidth]{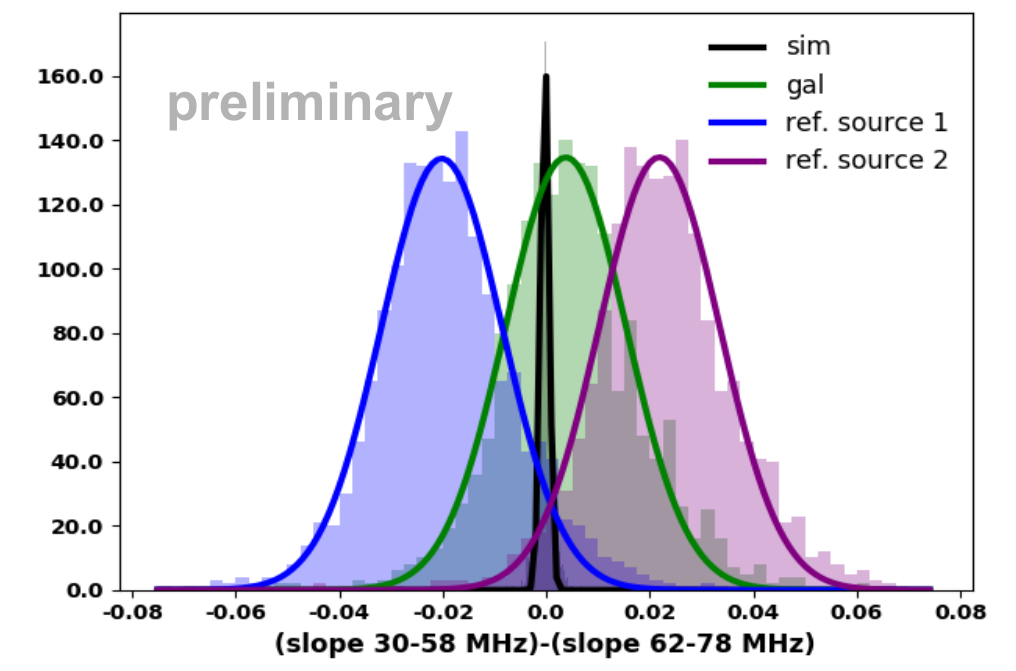}
	\caption{Difference of the slope of the frequency spectrum at high and low frequencies for LOFAR data using different calibrations and simulations.}
	\label{fig:slope_consitency}
\end{figure}

So far we used a calibration obtained from an in-situ measurement with a
reference source optimized for frequencies higher than our observation band.
The calibration with a reference source is consistent with a calibration using
the galaxy~\cite{Nelles2015a}. However, by modelling of the noise contributions
of the individual components of the signal processing chain we reduced the
uncertainty of the galaxy calibration to a value lower than the uncertainty in
calibration using the reference sources. The calibration constants obtained
with and without detailed modelling of the full signal chain methods are
displayed in Figure~\ref{fig:antenna_calibration}. In particular in the
frequency range between \SIrange{60}{80}{\mega\hertz} the uncertainty on the
galaxy calibration is now smaller than the systematic uncertainty arising from
two different calibrations of the reference source provided by the
manufacturer.
This reduces a frequency dependent bias of the measurement of the slope of the spectral shape of the signal as shown in~\ref{fig:slope_consitency}. This also enables detailed analysis of the frequency spectrum of the observed events that may be used as an additional observable sensitive to $X_{\max}$~\cite{Rossetto2017a}.

\section{Future Extensions}
Several extensions are currently planned, respectively already under construction, that will
increase the amount of data and extend the energy range for astrophysics
analysis with LOFAR.

\subsection{LORA Expansion}
In April 2018 we installed infrastructure for twenty more particle detectors at
five LOFAR stations close to the superterp. Installation of the detectors and
electronics is expected to be finished within 2018. These additional detectors
will increase the effective area for cosmic ray detection by approximately
40\% and enable the development of new trigger modes for specific classes of events. Of particular
interest are showers whose Cherenkov cone is located outside the superterp.
While the radio signal is greatly diminished in this geometry, it may be
possible to retrieve more information on the longitudinal shower development,
thus improving composition measurements and eventually constraining hadronic
interaction models.

\subsection{Advanced Trigger}
The trigger rate for cosmic ray detection is limited to approximately one
trigger per hour to not disturb astronomical observations of the LOFAR
telescope. This is achieved by requiring a signal over threshold in 13 out of 20 LORA
particle detectors. This yields virtually 100\% detection efficiency for showers above an
energy of \SI{0.1}{\EeV}, thereby defining the lower energy threshold for the
data useable for composition studies.
However, as a consequence of these trigger conditions most of the recorded showers are
of lower energies and approximately 80\% do not even contain an observable radio
pulse.

Reduction of the energy threshold for bias-free composition measurements while
keeping the trigger rate constant requires to focus on showers with a
detectable radio signal only. This is possible with a hybrid trigger that
requires a strong signal in the antennas in coincidence with a signal in the
particle array. Such an hybrid trigger will be implemented using a monitoring
channel implemented in the LOFAR system that records the voltage levels in the
antennas and can be used to send information to the LORA system.

Still, the current setup in LOFAR and all other radio experiments outside
Antarctica requires particle detectors to trigger on cosmic ray air showers.
Triggering on radio data only would instead allow for a very cost-effective increase
of the instrumented area. Therefore we are developing a self-trigger aiming for
a high reduction of RFI~\cite{Bonardi2017a}.

\subsection{LOFAR 2.0}
In the current configuration, LOFAR observations cannot overlap and data of HBA
and LBA cannot be recorded at the same time. As currently roughly 50\%
of the observations use HBA, and also data in-between observations
cannot be processed, the available time for cosmic ray observations is limited.
LOFAR 2.0 is a planned extension of the technical capabilities of LOFAR
allowing in particular for simultaneous observations with high-band and low-band
antennas and also to perform multiple observations independently in parallel.
This extension can thus significantly increase the available cosmic-ray data.

\section{Conclusion}
Precise measurements of the mass composition of cosmic rays are important to
understand their origin and propagation. LOFAR continues to measure the cosmic
ray mass composition with high precision in the expected transition region
between cosmic rays of galactic and extragalactic origin.  Several improvements
of the analysis chain are being implemented that  will allow for the reduction of the
systematic uncertainties on composition and energy measurement. Additional
upgrades are currently planned, respectively are already under construction, that will increase the
duty cycle  of LOFAR for cosmic ray measurements and reduce the energy
threshold for mass composition analysis.

\section*{Acknowledgements}
The LOFAR cosmic ray key science project acknowledges funding from an Advanced Grant
of the European Research Council (FP/2007-2013) / ERC Grant Agreement n. 227610. The project
has also received funding from the European Research Council (ERC) under the European Union's
Horizon 2020 research and innovation programme (grant agreement No 640130). We furthermore
acknowledge financial support from FOM, (FOM-project 12PR3041-3) and NWO (Top Grant 614-
001-454, and Spinoza Prize SPI 78-409). TW is supported by DFG grant WI 4946/1-1.
LOFAR, the Low Frequency Array designed and constructed by ASTRON, has facilities
in several countries, that are owned by various parties (each with their own funding sources),
and that are collectively operated by the International LOFAR Telescope foundation under a joint
scientific policy.
\section*{References}

\providecommand{\newblock}{}

\end{document}